\documentclass[]{aa} % for a referee version

\usepackage{graphics}

\begin{document}

  % \thesaurus{   % A&A Section 6: Form. struct. and evolut. of stars
   %           ()}

%
   \title{On the carriers of the celestial infrared vibrational bands and their excitation mechanisms }

\author{R. Papoular
\inst{1}
%          \and
%          C. Ptolemy\inst{2}\fnmsep\thanks{Just to show the usage
%          of the elements in the author field}
          }

   \offprints{R. Papoular}

   \institute{Service d'Astrophysique and Service de Chimie Moleculaire, CEA Saclay, 91191 Gif-s-Yvette, France\\
              e-mail: papoular@wanadoo.fr
%         \and
%             University of Alexandria, Department of Geography\\
%             email: c.ptolemy@hipparch.uheaven.space
%             \thanks{The university of heaven temporarily does not
%                     accept e-mails}
             }

   \date{ }
   \authorrunning {R. Papoular}
   \titlerunning {On the UIBs and their excitation mechanisms}
   \maketitle
\begin{abstract}

The recent influx of high-quality infrared spectroscopic data has prompted an extensive reassessment of various laboratory models in comparison with the observed UIBs (Unidentified Infrared Bands). As a result, significant modifications were brought to the original 
\linebreak
paradigms. The focus here is on the evolution of the coal model, characterized by 1) a shift towards less aromatic materials of the same family (kerogens), 2) the introduction of a new excitation mechanism (chemiluminescence), based on the capture of hydrogen atoms by carbonaceous dust. Both developments are intended to accomodate observations from a larger range of dust environments and evolutionary stages. This leads to a more quantitative description of dust composition and structure, and a better understanding of its history. In short, according to the present model, the evolution of dust from inception in the circumstellar shells of AGB stars, through strong interstellar radiation fields, to consumption in protostars, is approximately mimicked by progressively more aromatic materials, starting from the young and mostly aliphatic kerogens, through more and more mature coals, to the final stage of polycrystalline graphite. A similar family of materials is obtained in the laboratory by annealing in vacuum up to or beyond 3000 K. The composition, structure and IR spectrum of these materials are extensively documented in the geology literature.

The fundamental characters of this model, viz. chemical and structural disorder, and diversity of chemical bondings, naturally point to ways of further tailoring, in order to fit particular observations more closely.

%\vfill\eject

 %           }
 %
  \end{abstract}

%
%________________________________________________________________
\section{Introduction}
This paper is devoted to the understanding of the set of celestial IR (InfraRed) emission bands in the range 3-15 $\mu$m which have been loosely, but wittingly, dubbed Unidentified Infrared Bands (UIBs) well after their discovery in the '70s. Since then they have been thoroughly documented, using higher sensitivity and resolution. Particularly large amounts of high quality data were collected by the IR satellites ISO [1] and IRTS [2], which were launched and operated in the '90s. Knowledgeable reviews of these and other observations may be found, for instance, in [3,4,5].

While electronic transitions in atomic ions, such as Argon and Sulphur, are also detected in the relevant range, sometimes on top of a UIB, they are easily distinguished by their much narrower width. The only present consensus regarding the UIBs is that they are due to fluctuations of the electric dipole moment associated with the vibrations of the atoms of carbon-rich carriers. Different research groups do not agree as to whether the carriers are \emph{free-flying molecules} or \emph{solid-state grains.} Very closely related to this dichotomy is the issue of the excitation mechanism which gives rise to the IR emission. Roughly speaking, the free-flying molecule model has been associated from inception to \emph{stochastic heating}, i.e. random absorption of a single UV photon, followed by \emph{immediate thermalization} of the deposited energy, and subsequent cooling due to IR radiation; this of course requires very small carriers ($\leq$50 atoms) and strong UV irradiation. Moreover, these carriers are considered to be PAHs (Polycyclic Aromatic Hydrocarbons), which are planar clusters of hexagonal C rings with peripheral attached H atoms. No assignment to any known particular PAH has yet been proposed, but it is suggested that the problem will ultimately be solved by a size-distribution of neutral/ionized, more or less hydrogenated PAHs. In this model, separate particles, the Very Small Grains (VSG), have to be invoked in order to account for the observed underlying continuum, but no composition or structure has yet been specified for this family as well. Donn et al. [6] have discussed this model long ago, but Sellgren, Tokunaga, Boulanger and others [5,3,4] have recently spelled out some of the "puzzles" [5] presented by the new observations.

In this paper, I explore the extent to which these puzzles can be solved by a solid-state model [7], and more specifically the \emph{coal/kerogen} model. This differs from the PAH model on nearly every count. The composition, here, is not restricted to C and H, but also includes small amounts of O, N and S, as suggested by cosmic abundances. \emph{The structure is not planar or regular, but 3-D and highly disordered; it is partly aromatic, partly aliphatic and olefinic} depending on dust environment and history. To accomodate this diversity and disorder, the carrier size should be larger than a few hundred atoms. Most importantly, \emph{well defined natural terrestrial analog materials are proposed}: these are all members of the kerogen/coal "saga", which have been thoroughly documented during the past half century, and whose composition and structure have been abundantly described in the literature. The diversity of the celestial IR spectra, in relative feature intensities and in position and width, is matched by that of the terrestrial spectra, which itself is not erratic but known to be due to progressive release of S, N, O and H atoms upon ageing and/or heating of the material. No ionization or size distribution need be invoked. The structure accounts for the feature bandwiths and ensures the presence of a continuum where required. Finally, the excitation mechanisms invoked here are 1) steady state "thermal" heating for strongly illuminated dust, 2) chemiluminescence for weakly illuminated dust.

The coal model was first presented in 1989 [8]. At that time, modelling the available celestial spectra required the use of relatively highly aromatic coal analogs [9]. The data now available warrant the complementary use of the lesser aromatic analogs, kerogens [10]. These data also prompted the consideration and elaboration of a new excitation process [11]. I therefore believe that the coal/kerogen model is now in a position to cure the main riddles of the PAH model. These developments motivated the writing of the present report.

Section 2 is a short reminder of the main properties and behaviours of kerogens and coals, relevant to the present discussion. In Sec. 3, I delineate some of the main riddles signalled by authors who applied the PAH model to the interpretation of astronomical observations. In each case, I then discuss how the solid-state model can remedy the problem or provide an alternative interpretation. Section 4 discusses in more detail chemiluminescence as an alternative to stochastic heating. Operational conclusions are drawn in the last section.

\section{Some properties of coals and kerogens}

The following is a cursory summary of the properties of natural amorphous carbons of interest to the present issue. Details and abundant references to specialized literature can be found in [7,10].
\vskip1cm
2.1 \emph{Composition and Structure}

The term coal refers here to the residual part of the raw terrestrial material after minerals and free molecules preexisting in its pores have been disposed of by suitable, standard, physical and chemical procedures. \emph{This residue is almost exclusively composed of C, H and O, with traces of N and S.} While coal is extracted from deep in the earth, kerogen is obtained from the superficial sedimentary rocks after treatment by aqueous alkaline and other organic solvents. The solid residue can be considered as coal in ``dispersed" form (very small particles). \emph{It is in this form that most of the organic matter in meteoritic carbonaceous chondrites is found}, a useful hint for our present purposes.

\begin{figure}
\resizebox{\hsize}{!}{\includegraphics{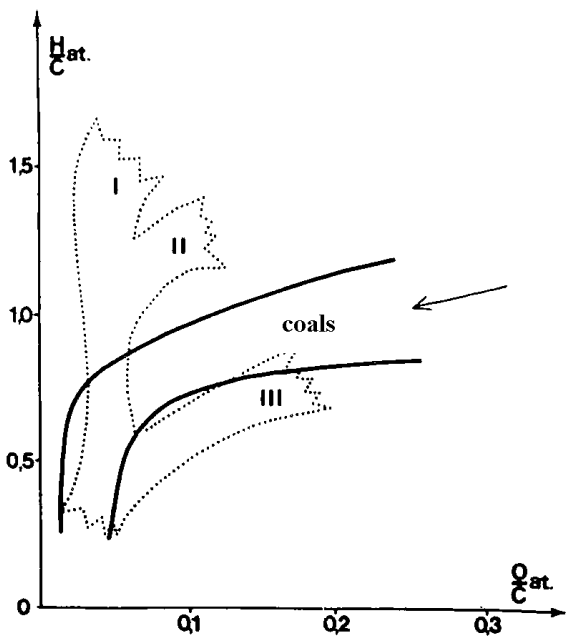}}
\caption[]{Van Krevelen diagram for coals and kerogens. The strip defined by the two continuous lines is the locus of coals of various origins and depths; I, II and III refer to kerogens of lake, marine and continental origins, respectively; the arrow shows the initial general direction of spontaneous changes in H and O atomic concentrations; see text and [7] for details.} 
%\label{fig1}
\end{figure}

The properties of coal do not depend heavily on the geographical origin of the material. As a result, each sample is concisely characterized by one point in the van Krevelen diagram (Fig. 1) with coordinates O/C and H/C (atomic ratios). The representative points of coals from different origins form an inverted-L-shaped strip. For coals extracted from various mining depths, but same geographical location, the strip nearly reduces to a line of the same shape; as the mining depth increases, the representative point follows the line down to the origin. The composition and properties of kerogens are more dispersed at high heteroatom concentrations, but ultimately merge with those of coals at low heteroatom concentrations. This graphical orderliness stems from the chemical properties of carbon.

Like in PAHs, the atoms in coals and kerogens are held together by $\sigma$ and $\pi$ covalent \emph{bonds}. But, while the former include only aromatic C \emph{sites}, the latter comprise all three types of sites: sp$^{1}$ (acetylenic), sp$^{2}$ (aromatic, olefinic) and sp$^{3}$ (aliphatic). This is a result of the presence of ``impurities" (O, N, ...) which impedes the tentency towards the more thermodynamically stable aromatic structures and leads, instead, to \emph{disordered materials}. As a consequence, PAHs know of only one type of C-H bond, while coals/kerogens admit of several others: C-H (non-aromatic), CH$_{2}$ (olefinic, aliphatic) and CH$_{3}$. Other important consequences are: a diversity of C-C bonds (single, double), the presence of  COH and C=O groups (which all have strong incidence on IR spectra), and -O- bridges which provide the structure with some flexibility and allow it to grow in 3-D.

Together with disorder comes the potentiality for \emph{evolution}. Figure 2 illustrates the displacements brought about, in van Krevelen's diagram, by spontaneous or induced losses of CO$_{2}$, H$_{2}$O and CH$_{4}$. In earth, these losses are stimulated by heat and pressure, which explains the trajectory of the representative point in Fig. 1 towards the origin as mining depth increases. The structural rearrangements which necessarily accompany these compositional changes are schematically outlined, in Fig. 2, along the left vertical axis, showing the aromatization trend along the evolutionary (maturation) track.

\begin{figure*}
\resizebox{\hsize}{!}{\includegraphics{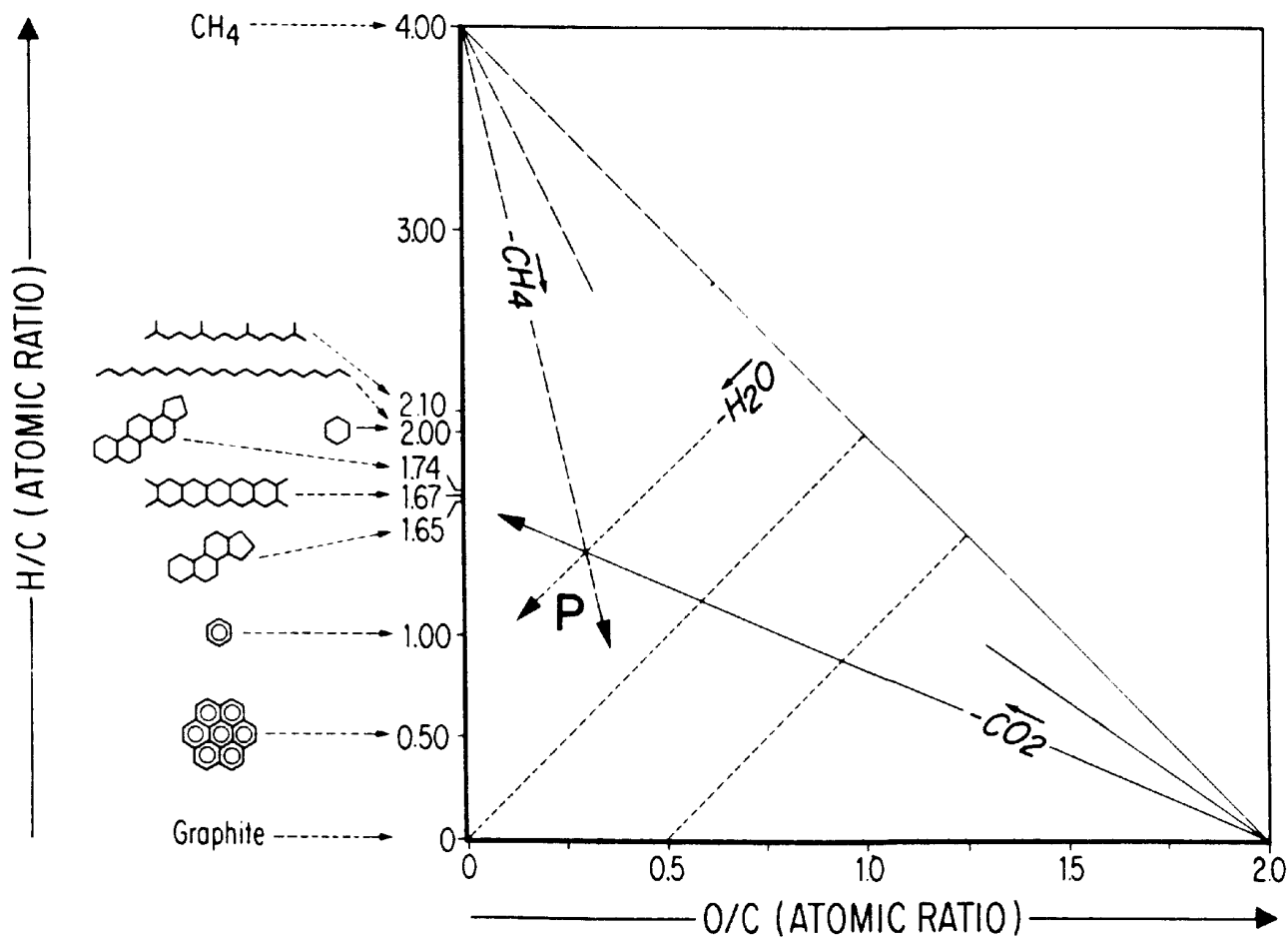}}
\caption[]{Van Krevelen diagram (after Durand [14], p122). Each arrow represents the direction of loss of a particular gas molecule in the course of evolution. A rough sketch of the main structural elements is given on the left side for a few evolutionary stages.} 
%\label{fig2}
\end{figure*}

In the laboratory, mild, prolonged, heating (annealing) of a low aromaticity sample prompts its representative point to quickly proceed along the same natural track it would have followed in earth, given time, had it not been extracted. This simple and coherent behaviour of coal/kerogen again emphasizes the tight link between composition and structure and provides a guiding line to find one's way in the \emph{diversity} of these materials.

Thousands of samples have been studied thoroughly all over the world, using all available analytical techniques: thermal, thermogravimetric, IR, Raman and NMR spectroscopies, etc. This made it possible to find correlations between spectral bands and assign nearly every band to a \emph{functional group} (a handful of associated atoms). These efforts culminated in the building up of representations of the structures of coals and kerogens as a function of their evolutionary stage, i.e. their location in the van Krevelen diagram (Fig. 1). The work of Behar and Vandenbroucke [12] is a good example of this development. They found that \emph{it is not possible to adequately describe the observed spontaneous and continuous changes in properties in terms of a limited number of small, specific, molecules (a conclusion to be extended below to the interpretation of UIBs)}. In order to fit the measurements on different samples, one has rather to build a \emph{random array\/} made of a dominant carbon skeleton with functional groups of heteroatoms (H, O, N, S) attached randomly to it, then statistically tailor the number of different bondings (C-C, C=C, etc), functional groups (C-H, C=O, etc) and aromatic or polyaromatic rings, etc. The variety of environments of the functional groups ensures that their characteristic vibrations will blend into bands of the right width. The relative intensities of the latter will change according to the concentrations of the corresponding functional groups. \emph{This scheme accounts for the continuous set of IR spectra observed along the evolutionary track\/}.

\begin{figure*}
\resizebox{\hsize}{!}{\includegraphics{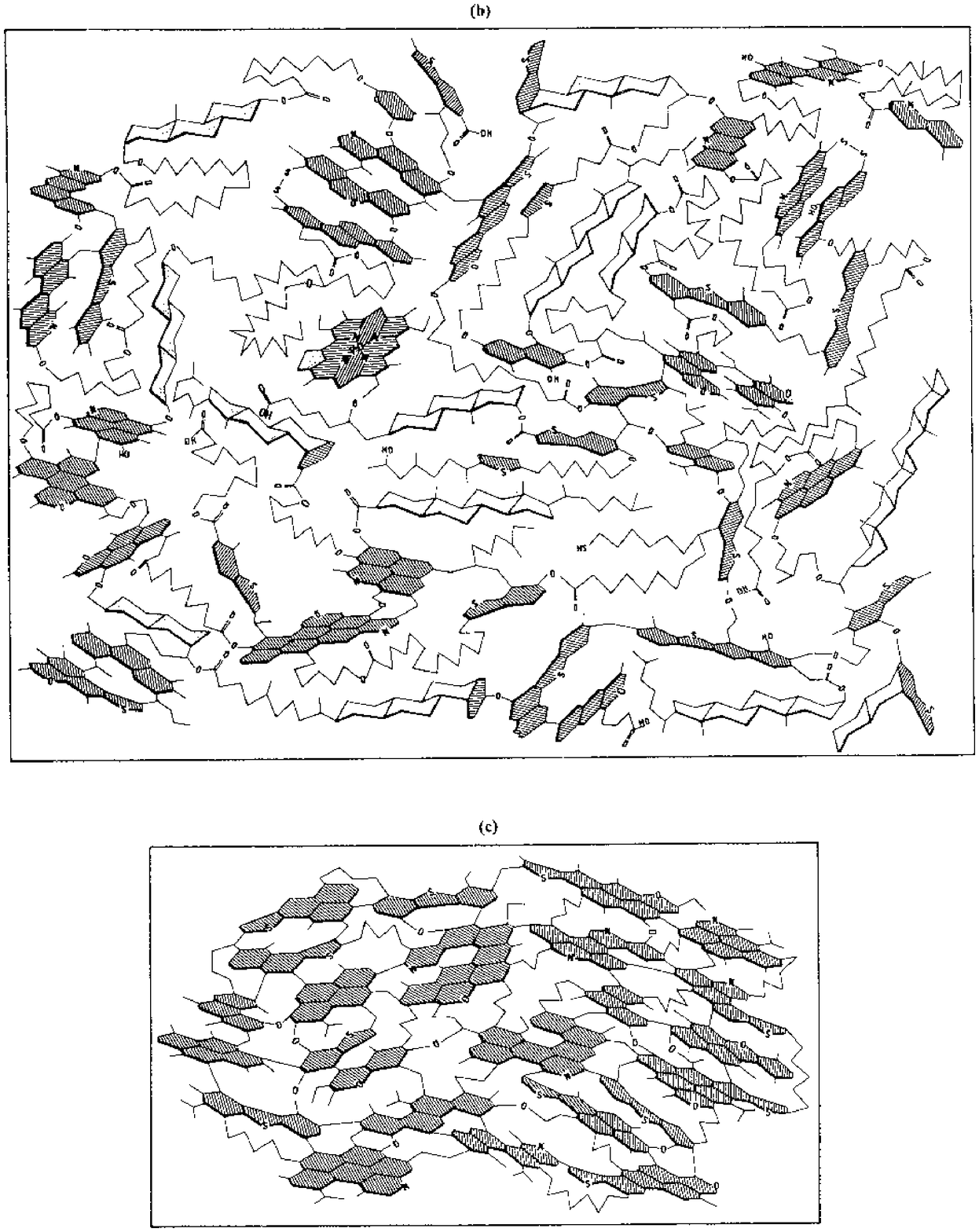}}
\caption[]{Chemical representation of a type II kerogen at an early (b), and a later (c) stage of aromatization (adapted from Behar and
Vandenbroucke [12]). Following common practice, C-H bonds are not represented. Aliphatic carbon chains (alkanes) are shown
as broken, undulating, lines. The aromatic clusters of benzenic rings are shaded. Various functional groups and oxygen
bridges are labeled. Note the increase in number and size of aromatic clusters, and the decrease in length of the aliphatic chains, from (b) to (c).} 
%\label{fig3}
\end{figure*}

Behar and Vandenbroucke give numbers for the structural parameters enumerated above and for 8 representative evolutionary stages. They also give sketches of the corresponding structures, examples of which are shown in Fig. 3. A remarkable feature in these sketches is that, even though the size and number of aromatic clusters are quite limited, all the main UIBs are present in the corresponding IR spectra, albeit with different intensities. The drawing of the clusters also highlights the 3-D nature of the structures. Note the large number of short aliphatic chains in the lesser aromatic samples; the aromatic clusters become large and dominant in the late stages of maturation.

\vskip1cm
2.2 \emph{The IR spectrum}

As a typical example, Fig. 4 (adapted from [13]) displays the IR absorbance spectrum of a young kerogen of type II (H/C=1.32, O/C=0.104). The peak wavenumber (cm$^{-1}$) and wavelength ($\mu$m), integrated intensity (cm/mg) and assignment of the bands are, respectively, (see [14]):

1. $\sim$3400($\sim$2.95); 46.2; OH stretch; due partly to chemically bonded OH groups and partly to adsorbed or trapped, not chemically bonded, H$_{2}$O molecules; the latter are responsible for the long redward tail, through H-bondings with other parts of the skeleton; peak position and band profile change considerably with H$_{2}$O content (depending on evolutionary stage or heat treatment). This feature is clearly distinct from the water ice band, which peaks near 3.1 $\mu$m and has a much steeper red wing (associated to that observed in the sky towards young stellar objects).  Adsorbed H$_{2}$O is probably absent from IS dust.

2. 3060 (3.27); C-H aromatic or olefinic stretch, barely visible here, and only measurable for very small H/C and O/C ratios.

3. 2920 (3.42); 48.2; blend of anti-symmetric and symmetric CH$_{3}$ stretch at 2962 and 2872, asymmetric and symmetric CH2 stretch at 2926 and 2853; CH stretch at 2890 cm$^{-1}$, respectively. Here, \emph{the non-aromatic C-H bonds reside mainly on non-aromatic structures, 
not at the periphery of PAHs like in the work of Wagner et al. [15].}

4. 1710 (5.85); 14; C=O (ketone) stretch.

5. 1630 (6.15); 11.4; disputed (but perhaps concurrent) assignments to H$_{2}$O deformation, quinonic C=O with H bond and C=C olefinic and aromatic stretch.

6. 1455 (6.87); 6.1; asymmetric CH$_{2}$ and CH$_{3}$ deformation.

7. 1375 (7.28); 1.2; symmetric CH$_{3}$  deformation.

8. 1800 to 900 (~5.5 to 11); 67.3; massif (underlying broad band) peaking near 8 $\mu$m, due to C...C and C-O stretch, C-H in-plane bend and OH deformation; in highly aromatic samples, there may be a contribution by electrons, in the form of a plasma resonance [16].

9. 930 to 700(~11 to 14); aromatic out-of-plane bending, depending on the number of adjacent protons; weakly contrasted 3 or 4 peaks, barely visible here; intensity  increases with mining depth but  never exceeds an intensity of 5.

10. An underlying continuum, roughly decreasing with frequency, is mostly visible in the near-IR, but also present farther to the red in all coals/kerogens; part of it is due to light scattering (in transmission measurements) and part of it to real absorption; it increases strongly with aromaticity.

 The intensity, K, given here (third number) is related to the absorbance, $\alpha$, by

\begin{equation}
$$\alpha$(cm$^{-1}$)=2.3 10$^{3}$ $\rho$ K/$\Delta$$\nu$$,
\end{equation}
where $\rho$ is the material density in g cm$^{-3}$ ($\sim$1) and $\Delta$$\nu$, the bandwidth 
in cm$^{-1}$. \emph{K (cm/mg) is $10^{-3}$ times the intensity usually defined in the astrophysical literature}; it is also related to the integrated cross-section per C atom, $\Sigma$, by
\begin{equation}
$$\Sigma$$(\mathrm{cm^{2} cm^{-1} C atom}^{-1})=4.7 $ $10^{-20} (1+\frac{H}{12 C}+\frac{16 O}{12 C})$ K$
\end{equation}

\hskip 3cm{$\approx4.7 $ to $ 7 $ $10^{-20}$ K.}

Not unexpectedly, a strong relationship holds between the bands at 2920 and 1455
 cm$^{-1}$ for all kerogen types and evolutionary stages: the ratio of their integrated intensities is $\sim$8. Also, K(3.4 $\mu$m) increases linearly with H/C, from 0 at $\leq$0.3 at least up to 80 cm/mg at H/C=1.3.

\begin{figure*}
\resizebox{\hsize}{!}{\includegraphics{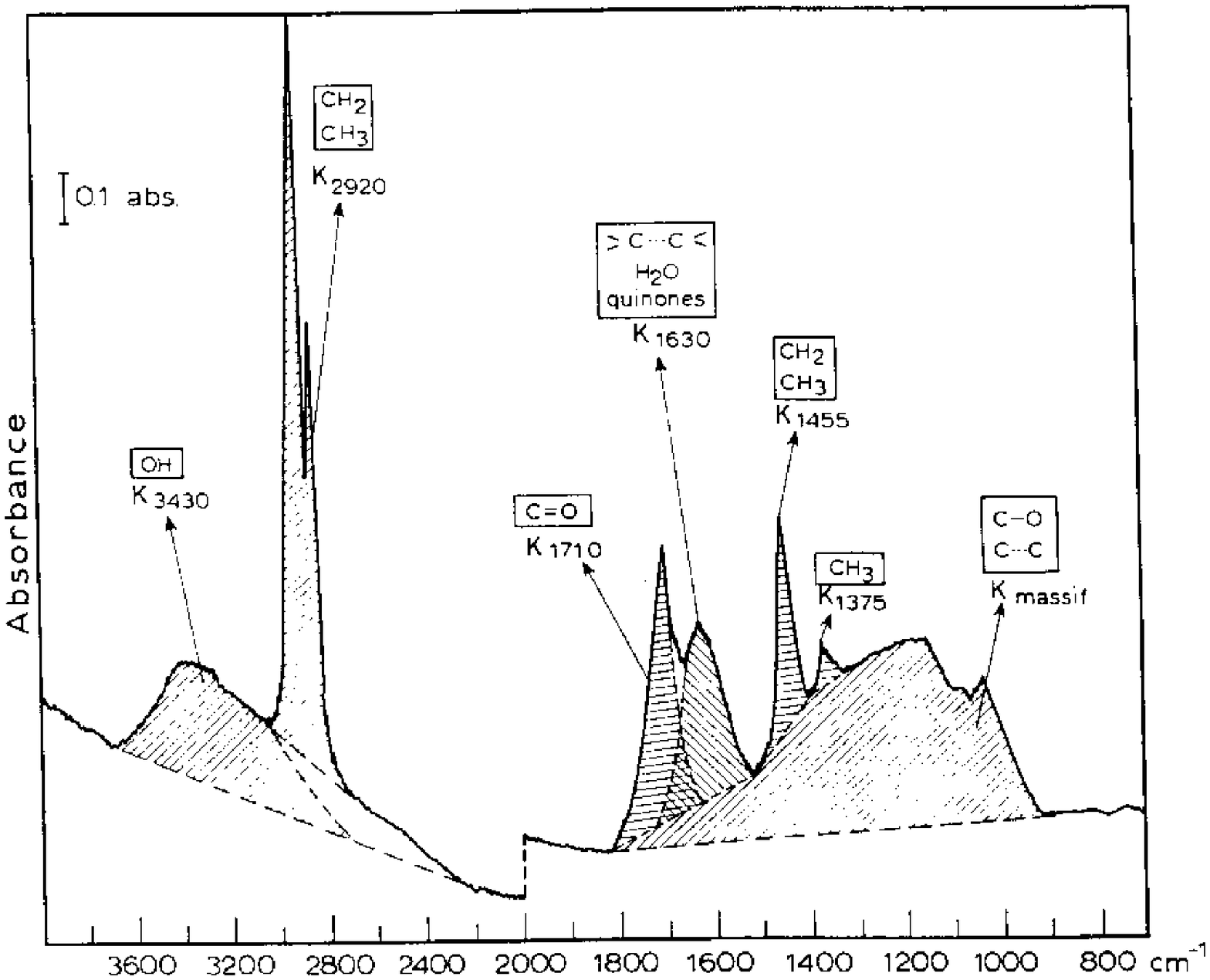}}
\caption[]{Typical IR absorbance spectrum of a type II kerogen, after Robin and Rouxhet [13]). The dashed areas
are used to determine the band intensities, K \emph{(see eq. 1)}. A number of partial assignments of bands are enclosed in squares; see
details in Sec. 3.} 
%\label{fig4}
\end{figure*}

It should be clear from the above that a single spectrum like Fig. 4 cannot, by far, convey the full diversity of coal and kerogen spectra. Series of spectra in restricted wavelength ranges and corresponding to successive degrees of maturation may help for this purpose. An example is given in Fig. 6 for the 3 $\mu$m range; others can be found in [10, 14].

\section{Puzzles and clues}

Here I discuss some of the problems encountered in applying the PAH model to the most recent astronomical observations. In enunciating these, I shall try to remain faithful to the wording of the referenced papers; this task is made easier by the authors' candour. The recent review by Sellgren [5] was particularly useful in this respect.
\vskip1cm
3.1 \emph{Carrier composition}

By definition, PAHs are pure hydrocarbons. However, the carriers of the UIBs are most likely formed in the winds of C-rich post AGB stars. In general, oxygen is not much less abundant  than carbon in these winds. Nitrogen and sulphur are also present in notable quantities. It is therefore difficult to understand why these three elements should be excluded from the dust formed in such environments. All the more so since the less abundant silicon shows up in the SiC feature at 11.3 $\mu$m, and, most importantly since the ISM (InterStellar Medium) harbors many families of molecules containing these so-called heteroatoms, especially O, mostly associated with C (see [17,18]).

On the other hand, minority atoms, mostly O, are present in coals and kerogens in known concentrations, and these inclusions have important consequences: 1) they considerably enhance the dipole moment of some vibrational transitions involving C atoms, such as the 6.2 $\mu$m stretch of the C=C bond; 2) they give rise to features observed in the sky in the mid-IR, such as the band at 1710 cm$^{-1}$ (5.85 $\mu$m) and the massif at 7-9 $\mu$m  [10], as well as in the far IR, such as the 30 $\mu$m blend associated with COH wagging (see [19]); 3) minority atoms efficiently interfere with the tendency of C atoms to settle into an ordered aromatic structure; \emph{hence the amorphous 3-D structure of Fig. 3, which determines all the favourable properties of this model}, as shown below.
\vskip1cm
3.2 \emph{Feature bandwidths}

The UIBs are strikingly wider than the bands of any observed and identified molecule, the narrowest width being about 15 cm$^{-1}$. Even at the highest resolution, no rotation-vibration decomposition is discernible, although such decomposition was clearly demonstrated, with the same spectrometers, for many identified molecules, such as H$_{2}$ and H$_{2}$O. Neither is it possible to assign these widths to lifetime broadening, since the typical width of 30 cm$^{-1}$ would correspond to a lifetime of about 0.1 ps, way shorter than the radiative lifetime ($\sim$0.1 s) and non-radiative lifetimes, ranging between 1 and 1000 ps. Thermal Doppler effects are also excluded for the masses and temperatures of particles of interest. Thus, although UIBs can be fit by one or more Lorentzians or Gaussians, these are too wide to carry physical meaning: the UIBs are definitely broader than laboratory PAH features \emph{in absorption} (see [20], [5]).

 In \emph{emission}, highly excited molecules may have much broader features than in absorption because of \emph{anharmonicity and spectral congestion}. This is illustrated by Williams and Leone (21], who excited gas-phase naphtalene molecules (plane clusters of 2 benzenic rings, 18 atoms) with short laser pulses of photons of energy 5.3 and 6.6 eV. Monitoring the outgoing radiation at 3.3 $\mu$m, (C-H stretch), they found its width to decrease from $\sim$100 to $\sim$30 cm$^{-1}$, while the peak shifted to the blue by 45 cm$^{-1}$, as the energy content of the molecule was reduced by on-going radiative emission. No such behaviour is observed in the sky.

Being associated by birth with single-photon excitation (stochastic heating), the PAH model is therefore faced with a dilemma: either the molecule is so small, and the photon so energetic, that the "temperature" can reach about 1000 K and the emission is strong enough as to be detected, but then with unacceptably large widths and red shifts; or the molecule is large and the photon energy low so that no undesirable broadening and/or red shift occurs, but then with unacceptably weak intensities and narrow widths. Recent attempts to solve this dilemma relied on the strong but unsupported assumption that the band position and widths of the features are the same functions of temperature for all interstellar PAH sizes, or on very specific mixtures of selected PAH cations supposed to be independent of history and environment (see discussions by Verstraete et al. [22, 23] and Boulanger et al. [24]).

By contrast, the \emph{disordered solid-state} model assigns the features to the synchronous vibrations of specific functional groups, and most of their width to the dispersion of similar functional groups among various environments, a situation corresponding to small-scale inhomogeneity [20]. The widths of UIBs is indeed typical of solid, disordered materials [25] and, in particular, coals and kerogens [10]. The disorder, here, is mainly induced by heteroatoms and is, therefore, chemical and structural in nature (Sec. 2). Within limits, and especially below 500 K, feature widths are not sensitive to temperature or grain size, provided it is large enough (a few hundred atoms) to accomodate the required diversity of environments for a given average chemical composition. Under such circumstances, small degrees of ionization and dehydrogenation are not likely to alter the spectrum notably, in stark contrast with PAH behaviour.

While this explains the remarkable stability of the UIB widths [3,22], it also leaves room for the observed diversity [20], which may result from different compositions at inception, as well as different thermal and radiative processing in the interstellar medium. Because of this disordered, solid-state structure, the coal/kerogen model also adequately mimicks the celestial massifs at 7-9 $\mu$m (see fig. 4) and 11-13 $\mu$m [9], which are major puzzles for the PAH model [5].
\vskip1cm
3.3 \emph{Carrier aromaticity}

The use of the term PAH implies full aromaticity. This radical assumption was initially suggested by the strength and ubiquity of the 3.3, 6.2 and 11.3 $\mu$m features. However, it was not supported by later observations illustrated by fig. 5 for the stretch bands and in [7] for other ones: distinct emission features between 3.3 and 3.6 $\mu$m are seen towards AGB stars such as IRAS 05341+0852, 04296+3429 and CRL 2688, in Planetary Nebulae (e.g. Hb 5 [26]) and towards at least one Nova: Nova Cen 1986 [27]. In some cases, they are even stronger than the 3.3 $\mu$m feature, as in absorption towards the Galactic Center, where the aromatic feature is dwarfed by a massif with peaks near 3.4 and 3.5 $\mu$m.

\begin{figure}
\resizebox{\hsize}{!}{\includegraphics{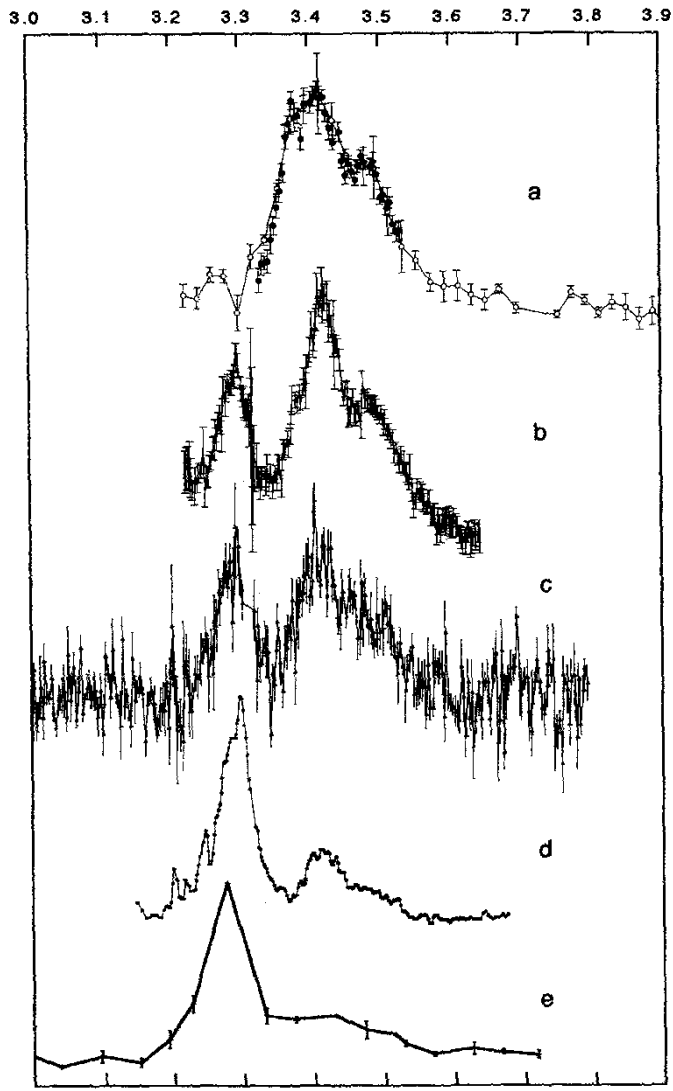}}
\caption[]{The C-H stretch band of a) GC/IRS6E, in absorption (Pendleton et al. [28]); b), c), d)
3 post-AGB stars in emission: IRAS 08341+0852 (Joblin et al. [29]), IRAS 04296+3429 (Geballe et al.
[30]) and CRL2688 (Geballe et al. [30]); e) reflexion nebula NGC 2023 in emission (Sellgren [31]).} 
%\label{fig5}
\end{figure}

\begin{figure}
\resizebox{8.4cm}{15.2cm}{\includegraphics{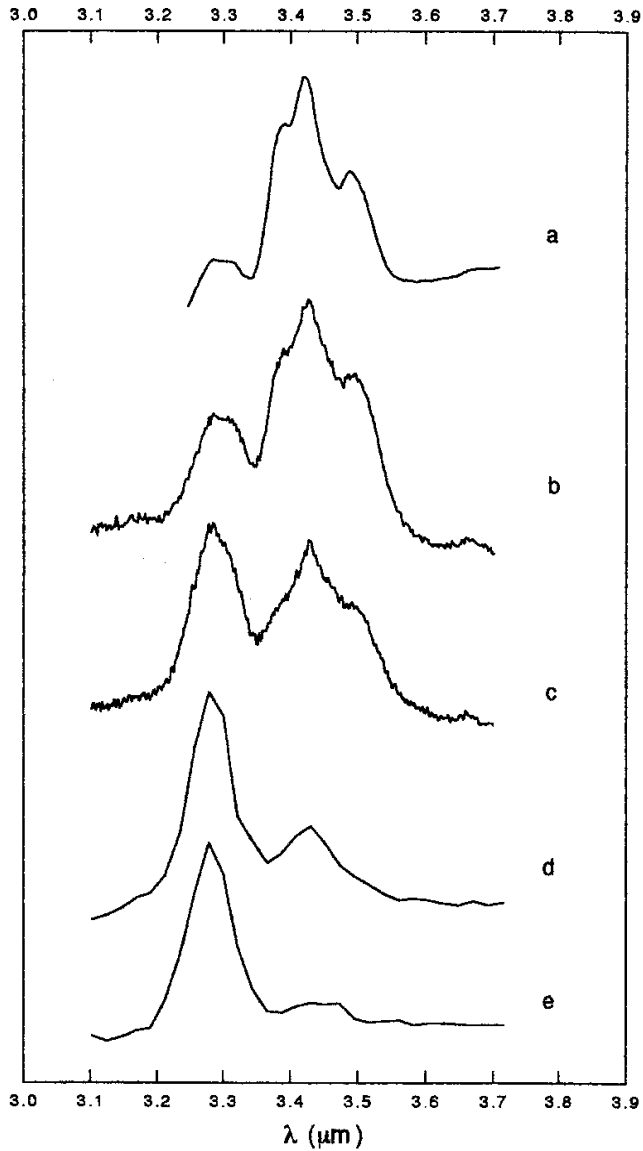}}
\caption[]{The C-H stretch absorption band of coal as a function of evolutionary stage: a) Vouters mine
(O/C=0.06, H/C=0.75), b) Mericourt mine (O/C=0.028, H/C=0.59), c) Escarpelle mine (O/C=0.018, H/C=0.46), d) and e)
Escarpelle sample annealed at 525 and 600 K, respectively. Adapted from Guillois [32]).} 
%\label{fig6}
\end{figure}

 The initial assignment of these "red" features to hot bands (v$>$1) of the C-H str transition has proven unconvincing because of their strength and the non-detection of the correlative first harmonic to be expected near 1.67 $\mu$m [15, 26]. It seems therefore more reasonable to assign them to the well documented aliphatic and olefinic, symmetric and asymmetric, C-H stretches at 3.4, 3.46, 3.515 and 3.56 $\mu$m, in CH, CH$_{2}$, CH$_{3}$ and aldehydes (-CHO) groups [33,34]. Figure 6 illustrates this statement by setting, side by side with the spectra of fig. 5, the spectra of coals at increasingly higher stages of aromatization (see Sec. 2): not only is the fit of corresponding spectra remarkable, but the regular decrease of the red blend together with the rise of the 3.3 feature confirms their assignment to non-aromatic vibrations.

Neither is it obvious that the 6.2 and 11.3 $\mu$m features are characteristic of aromatics. Indeed, while the 11-13 $\mu$m massif with 3 or 4 peaks, observed towards post-AGB stars may reasonably be ascribed to the out-of-plane vibrations of aromatic C-H bonds [9], it has proven very difficult to devise a PAH cluster that could display a highly contrasted, strong and lonely 11.3 feature as observed in most IS (InterStellar) spectra. According to [35], this requires "rectangular" PAHs with 100 to 150 C atoms, a very special requirement indeed! (see next subsection). Moreover, the expected correlation between the 3.3 and 11.3 features is not confirmed by observations [47, 52, 63]. Even the exact spatial coincidence of the intensity peak of the 11.3 feature with that of the 3.3 feature appears to be in doubt (see [63, 64]).

 On the other hand, a (more familiar) alternative or concurrent assignment for the 11.3 $\mu$m band could be the wagging of 
the two H atoms of a vinylidine group (R$_{2}$-C=C-H$_{2}$), which also occurs at the desired frequency [33]. Clearly, the constraint of full aromaticity is not warranted: an understanding of the ``3.4" feature, as well as of the diversity of UIB spectra, is better served by thinking in terms of \emph{evolution of IS dust composition and structure}, both intimately connected and induced by prolonged heating (Sec. 2) and irradiation [36]. While this is natural, straightforword and quantitative in the coal/kerogen model, it seems to be hardly possible with compact PAHs, whatever their size and degree of ionization or hydrogenation.

\vskip1cm
3.4 \emph{Spectral continuum}

Known PAHs, even when ionized, cannot provide enough visible \emph{continuum} absorption to account for the energy in the emission features observed around UV-poor reflection nebulae [5].Neither can they mimick the unpolarized near-IR continuum detected towards the same objects [5,3]. This, of course, does not come as a surprise since free-flying molecules are not usually known for their continuum, although they do exhibit lines or bands in the vis/UV. Herein lies yet another, perhaps the main, reasons for invoking PAHs much larger than initially envisioned [5,35]. Unfortunately, none has yet been produced in the laboratory, if this is possible at all. Indeed it is well known that, in the absence of externally applied pressure, a planar sheet of compact benzene rings (graphene) of increasing size tends to bend, warp and finally folds upon itself to form a fullerene or a carbon nanotube[37].

Now, these particular carbon structures indeed have features in the UV/vis but they also have features elsewhere, such as the four conspicuous IR bands of fullerene at 526, 576, 1183 and 1428 cm$^{-1}$ [38]. None of these has been detected in the sky despite intensive seach.

Moreover, even these structures do not display mid- and far-IR continua, which points to still another puzzle of the PAH model: there is also a \emph{mid-IR} continuum underneath the UIBs; for want of a better assignment, this has been attributed to a separate dust component, the very small grains (VSGs), but its composition and structure could not be specified further [5, 39]. Note that they were thought to be small solely on the grounds of the self-imposed constraint that their emission should be excited upon absorption of a single UV/vis photon. Recent surveys of the spatial variations of this continuum have added to the confusion and seem to require again very special behaviour of the assumed separate component in order to match the systematic changes in the ratio of features to continuum [24, 39, 40, 41, 42].

Of course, theorists never lack an explanation, but all tentative applications of the PAH model seem to converge towards particles much larger than invoked in the first place [5, 24, 35, 43, 44]. At that point, it must be recalled that \emph{bulk} carbonaceous materials indeed display a continuum, but the profile and intensity of the latter depend heavily on the composition and structure of the material [7]. Thus, the continuum of glassy carbon is very strong and varies approximately as $\lambda$$^{-1}$ at least as far as 100 $\mu$m, but coals and a-C:H (amorphous hydrogenated carbon) are semi-conductors whose absorption falls down sharply towards the red at a cut-off wavelength ($\sim$1 $\mu$m) which increases with aromaticity. The lower aromaticity kerogen has a cut-off at shorter wavelengths and, hence, a weaker mid-IR continuum, better suited to the high contrast of the UIBs; it has also a near-IR continuum rising towards short wavelengths;  it has the further advantage of carrying both continuum and bands [10], thus eschewing the need for two different carriers.
\vskip1cm
3.5 \emph{Near- and far-UV bands}

Simple aromatic molecules are known for their near- and far-UV transitions between ground and excited electronic states[45]. None of these was detected in the sky, either in absorption or in emission (fluorescence) [6], although several attempts were made to identify the DIBs (Diffuse Interstellar Bands) with such bands. In this field, too, hopes now hinge upon large 3-D systems [46].

Indeed, it was shown that high-temperature treatment (HTT) of coals favored aromatization and the growth of a \emph{single} near-UV feature which irreversibly becomes narrower and bluer [7]. In some favourable cases, the end point of this evolution is \emph{polycrystalline (disordered) 
\linebreak
graphite}, with a single, conspicuous, feature near 217 nm, on top of a continuum which increases sharply to the blue and decreases slowly to the red [7]. Then, of course, the material can hardly be designated as a PAH.
\vskip1cm
3.6 \emph{Infrared bands}

Regarding wavelength and bandwidth, the UIBs are remarkably stable against large variations in the environment and irradiation. This stability is incompatible with an assignment of the bands to a soup of free-flying independent PAHs, neutral or ionized and more or less hydrogenated, as suggested by a number of authors in the past, since later laboratory studies showed that individual PAHs are highly sensitive to irradiation wavelength and intensity (see discussions in [5, 24, 48, 49, 50]. In trying to use these parameters to fit observations, one is often led to conflicting or paradoxical conclusions (see discussions in [22, 23,35, 49, 50, 51]).

One is therefore led again to prefer a solid-state (bulk) model: in that case, ionization and dehydrogenation can only affect the surface of the grain, a minor fraction of the whole. Also, the dust is then much more robust against destruction by shocks. Moreover, the structure and composition of a bulk, amorphous, material as coal/kerogen can \emph{evolve slowly, spontaneously or upon heat treatment} as shown in Sec. 2. This is accompanied by subtle changes in band wavelengths and widths (small fractions of a micrometer), also observed in the sky (see [10, 20]). Very large changes in relative intensities are also observed, especially in the ratios [3.3]/[3.4], [3.3]/[11.3] and [11.3]/[12.7] (see[47, 52]). On terrestrial analogs, most of these can be explained by coherent and systematic changes of composition and structure brought about by heat or slow spontaneous evolution. The most spectacular and best documented case is that of C-H stretch (see fig. 6): Dischler analyzed in detail the evolution of each component of this group of vibrations [34] .

Thus \emph{the spectral diversity of coals/kerogens of different degrees of maturation (ranks) and origins matches the diversity of astronomical spectra. The main astronomical features associated with carbonacious IS dust, including the continuum, are provided, to various extents, by coals/kerogens of various ranks. Different coals/kerogens or mixtures thereof can account, roughly at least, for astronomical spectra from different IS environments. These models introduce no significant spurious feature anywhere in the explored spectrum.} Examples of quantitative fits between UIBs and models can be found in [7] for the UV/vis range, here in Fig. 5 and 6 for the near-IR, in [10, 11] for the 5-10 $\mu$m range, in [9] for the 11-13 $\mu$m range and [19] for the far-IR range; see also [32].

That is not to mean that a perfect match between celestial and laboratory spectra is easily obtained. For one thing, the high quality of ISO spectra has revealed many new weak features, or details of the profiles of previously known features, which have yet to be elucidated (see [35, 52]). More importantly, no satisfactory interpretation is yet available for the 12.7 feature, nor for the 11.3 feature being usually so much stronger than the underlying blend (11-13 $\mu$m) which is commonly assigned to C-H mono, duo, trio and quarto out-of-plane bends (see Sec. 2). \emph{The 11.3 and 12.7 bands are not present with a high contrast in the solid- state model spectra.}

But the coal/kerogen model provides at least a framework with definite guidelines and valuable assets:

-representative natural samples are available with known composition, structure, physical properties and band assignments;

-well-defined heat treatments can alter these properties controllably and coherently;

-the range of possible model spectra can be extended by using samples of different depths and geographical origins, thus scanning the whole van Krevelen diagram (fig. 1), i.e. a large range of \emph{natural} structures. 

-the existence, in space, of carbon chains and molecules as large as acetone  is easier to understand in terms of breakup of large, inhomogeneous, kerogen-like, grains 
\linebreak
(formed in ``dense" circumstellar shells and destroyed in strong IS shocks), rather than in terms of synthesis in the tenuous IS medium.
\vskip1cm
3.7 \emph{Excitation}

The relatively high ratio of short to long wavelength UIB intensities initially led to the conclusion that the emitter temperature should be very high. Under the assumption of stochastic heating by absorption of a photon, and for the relatively small PAHs initially considered, calculations showed that such a temperature already requires the absorption of UV to far UV photons [53]. The steady increase of the absolute intensities of the UIBs with ambient UV radiation field intensity (G value) over 4 to 5 orders of magnitude has been considered as the strongest argument in favour of stochastic heating by UV photons [4]. These conclusions are in conflict with a number of astronomical and laboratory facts.

1)The relative UIB intensities from reflection nebulae are independent of the temperature of the illuminating star, i.e. of the hardness of the radiation field, from 22000 down to about 5000 K (see [5]); this is at variance with the sharp decrease of dust temperature with photon energy predicted by calculations [53].

2)Shortward of 10$\mu$m, the shape of UIB spectra of several galaxies that have been studied are strikingly similar for nuclear regions, active regions in the disks, as well as quiet regions [49]. A similar stability is observed among spectra of regions of hugely different G values, which are accompanied by large variations of star temperature and, hence, of the spectrum of the exciting radiation. Again, one should therefore expect notable non-linear variations of relative band intensities, which is not the case [4]. the authors of [49] conclude that the global relation between UIB intensity and UV illumination might be only indirect. This is corroborated by the observation of UIB emission from poor-UV regions, such as galaxy M31 [61].

3)The ``small PAH-stochastic heating" paradigm is also confronted with the stability of the UIB wavelengths and widths against environmental variations, especially in view of attendent dehydrogenation,ionization or downright destruction of the molecule [4, 5, 24, 39, 48, 50] (see Sec. 3.2).

4)It appears from the previous subsections that all available spectral evidence weighs in favour of bulk, solid-state models of IS dust, as opposed to free-flying individual molecules. Recent discussions involving PAHs also invoke 
large ``molecules", up to 1000 atoms or more [5, 24, 35, 43]. Now, according to the former calculations, this leads to prohibitively low dust temperatures and undetectable UIB emission. One way out of this paradox is to hypothesize very special, but as yet not measured, electronic properties: high photon absorption cross-sections, absorption cut-off sliding to the red as the particle size increases, etc. This becomes still more accrobatic if one is to account for UIBs detected in poor-UV environments (see [22, 54]). 

The coal model is confronted with a symmetrical puzzle as long as it holds to the simplest excitation process: thermal equilibrium. This is only effective in strong radiation fields, in the immediate vicinity of hot stars, as is the case with post-AGB stars and PPNe (Pre-Planetary Nebulae); see [9]); but it is out of the question in RNe (Reflection Nebulae) and questionable in other instances. All models are therefore confronted with the task of conceiving a third excitation mechanism.

A significant hint towards this goal is provided by the spatial distribution of the intensity of the 3.3 feature across the Orion Bar observed edge-on [55]: it forms a narrow peak between the ionization front (delineated by the Br $\alpha$ and P$\alpha$ lines) and the H$_{2}$ excitation peak (delineated by its ro-vibration transitions in the ground state). While the sharp drop towards the ionization front may perhaps be attributed to partial or total destruction of the carrier in the (hot) HII region, the drop on the opposite side is more problematic because it cannot be due to a decrease in the column density. Sellgren et al. [55] finally assigned it to the extinction of the stellar radiation  by molecular hydrogen, assuming a density of 10$^{4}$ cm$^{-3}$ for the latter. However, if this was the case (with a small PAH, as required by single-photon heating), one would also expect a change in the spectrum of the exciting radiation and, consequently, a change in the 3.3 emission feature across the Bar. What is observed, in fact, is a remarkable stability of wavelength, width, [3.4]/[3.3] and [continuum]/[3.3] ratios.

Now, the steep, large and opposite intensity variations of the ionic and molecular hydrogen spectral features observed to occur across the Bar suggest that, in between the ionic and molecular regions, their must be a narrow transition layer of atomic hydrogen, roughly coincident with the 3.3 feature peak. A similar situation has been described in NGC 7027 [56] and seems to have been observed at the edge of high-latitude cirruses [41, 65]. This led us to surmise that the excitation of the UIBs is linked with the presence of these atoms, rather than directly with the ionizing radiation [57]. Such a scenario (developed in the next section) naturally and straightforwardly explains the small transversal extent of the 3.3 layer, without having recourse to uncertain or qualitative assumptions as to the destruction of dust carriers towards the exciting star [41], or steep extinction of the UV towards the molecular cloud.

\section{Chemiluminescence}

Hydrogen \emph{radicals} (free atoms) are known to be highly reactive, especially with hydrogen and carbon. When such a radical hits a hydrogeneted surface, it is likely to form, with an H atom of the surface, a H$_{2}$ molecule, which will promptly escape, leaving a \emph{dangling bond}. This catalytic reaction between two H atoms is the main route envisioned by astrochemists to explain the high rate of formation of molecular hydrogen in space: here, the solid surface provides the third body which is necessary to ensure conservation of momentum. Since carbon is known for its high catalytic efficiency, our IS carbon dust is a good candidate for the job, provided it can present a large fraction of dangling bonds per C atom. It must therefore be disordered and porous, as are coals and kerogens. Note that the nascent H$_{2}$ molecule carries away a large part of the energy made available by this exothermic reaction; it must therefore be ro-vibrationally excited and this may be the source of part of the H$_{2}$$^{*}$ lines observed in PDR's and at the edge of molecular clouds, between ionized H and molecular gas (CO).

Now, when another H atom approaches the dangling bond left on the dust, it is very likely to be captured and form a new strong C-H bond. The chemical (potential) energy of this bond, $\sim$4.5 eV, is insufficient to dissociate other bonds or excite higher electronic states; it is then available as ground-state vibrational energy for the atoms in the grain. Initially, this energy is essentially in the form of C-H stretch. However, it quickly spreads around, exciting the other characteristic (normal) modes of the given particle. The bond anharmonicity (non-linearity of restoring forces) opens channels for the available energy to be shared among the modes, which otherwise would be isolated. This process is called IVR (Internal Vibrational Redistribution), and is usually very short-lived, of order 1 ps, after which the energy deposited in an isolated grain can only escape through IR radiation, with a time constant of order 0.1 s.

Such \emph{infrared chemiluminescence}, following molecular reaction in the gas phase was observed long ago [58] and often thereafter. Similar processes occur in an isolated gaseous molecule upon absorption of a vis/UV photon, as in the experiment of Williams and Leone [21], referred to above. The problem of theoretically predicting the chemiluminescence spectrum is extremely difficult to solve because of lack of detailed knowledge of the various couplings  (force fields) between atoms in a given molecule: electrostatic dipole and higher order moments, van der Waals and other forces, hydrogen bonding, etc.This is compounded by the length of dynamical calculations in the case of large systems, such as those that are now envisioned as dust models. Fortunately, commercial software is now available which is dedicated to numerical simulation of molecular structure, dynamics and reactions. I have used such gear to collect quantitative, semi-quantitative and qualitative data on the structure of coal/kerogen model particles as large as 500 atoms, the interaction of H atoms with such particles and the resulting vibrational spectra [59, 60].

The main result of these computations, of interest for present purposes, is that, to a large extent, \emph{ the predicted emission spectrum is very similar to the absorption spectrum, independent of particle size and colour temperature of the illuminating star. No immediate thermalization occurs}. Moreover, since the only relaxation route is through IR radiation, \emph{the luminescence efficiency in the UIBs should be very high} (of order 0.1 if it is defined as the ratio of IR energy emitted over deposited energy).

Chemiluminescence thus solves the puzzles encountered above with the two other excitation mechanisms, \emph{provided enough atomic H is available}. If there is a luminous star near by a molecular cloud, it will give rise, at the edge of the cloud, to a PDR (PhotoDissociation R
egion) such as the Orion Bar (referred to in Sec. 3.7), with contiguous ionized, atomic and molecular regions, successively in the radial direction from the star. The \emph{quantity} of chemically excited dust in the atomic layer, and hence the UIB intensities, will obviously depend on G, but the IR emission spectrum will not! This is all compatible with the roughly linear increase of the individual UIB intensities versus G, as well as with the stability of their spectrum over 5 orders of magnitude of G [24]. Moreover, because the rates of dissociation of H$_{2}$ molecules and ionization of H atoms depend on the flux of UV photons, the UIB intensities will depend, in absolute but not relative value, on the stellar colour temperature, which is not the same for all G's. This, together with the random variations of molecular density at the cloud edge, accounts for the observed excursions from strict linearity of intensity with G [24].

Of course, this does not exclude the coexistence of thermal equilibrium heating of grains if the ambient radiation field is strong enough. In the opposite case of no neighbouring luminous star, atomic H is still present, but in smaller quantities, if only for the diffuse galactic IS radiation field; this accounts for the UIB luminosity of nebulae at high Galactic latitudes, such as the Chameleon cloud or the Ursa Major cirrus [41]. 

A further observational support for this excitation model is the spatial coincidence of the UIB emission with the
\linebreak
``filament" of excited H$_{2}$ emission at the edges of NGC 7027 [56] and of the $\rho$Oph main cloud [65]: indeed, as indicated at the beginning of this Section, chemiluminescence induced by atomic hydrogen is genetically associated with the production of hydrogen molecules which carry away part of the chemical energy of the incoming H atom and are, therefore, ro-vibrationally excited.

In application of these ideas, a chemical kinetic model was built [11] for the galaxy M31, whose Diffise ISM emits the UIBs although it is nearly devoid of far UV radiation [61]. A good numerical fit to the observed IR intensities was obtained assuming an atomic H density between 50 and 200 cm$^{-3}$ (in agreement with the range determined for the edge of a typical molecular cloud in our Galaxy) and a chemiluminescence efficiency of 0.3, in agreement with the numerical chemical simulation referred to above.
\begin{figure}
%\vspace{8cm}
%\begin{center}
%\scalebox{0.5}[0.5]{
\resizebox{\hsize}{!}{\includegraphics{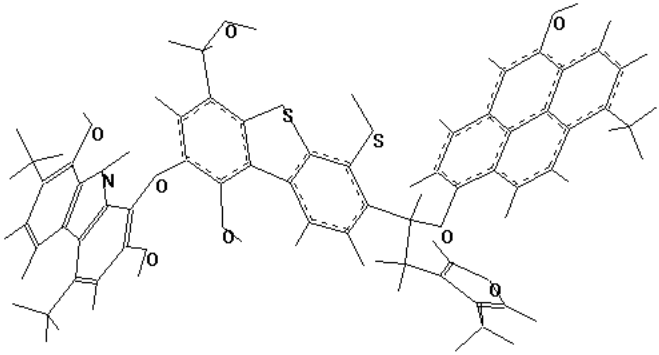}}
%\end{center}
\caption[ ]{The optimized structure of the building block (101 atoms) of the reference macromolecule (493 atoms) modelled in [60].}
\end{figure}

The essential difference between this excitation mechanism and stochastic heating does not lie in the energy deposition process (atomic H capture or photon absorption), but in the unwarranted assumption at the core of the stochastic heating theory, that \emph{thermalization of the absorbed energy occurs before radiative relaxation}, only because the latter may require up to 1 s. Common wisdom mixes up IVR with thermalization and the last decades of the past century were necessary to discriminate between various types of \emph{chaos} and between various degrees of \emph{randomization} (see bibliography in [60]). In momentum-configuration phase space, thermal (statistical) equilibrium is the extreme case where all points of the energetically available phase space are equally probable and ultimately visited by the representative point of the molecular system. This stage is only ideal and, in fact, very difficult to reach on a microscopic scale and in the finite time of experiments, but it is easier to approach when dealing with macroscopic assemblies of identical particles interacting with each other and with a ``thermal bath" (itself an ideal concept!). By contrast, experiments resolved in time,  space and wavelength have shown that the vibrations of atoms in an isolated system may retain a high degree of coherence for a long time. \emph{That means that only parts of phase space are visited, that randomization is not complete and that a temperature cannot be defined}. I have shown numerically that this is the case for a disordered system, even as large as 500 atoms [60]. For this purpose, I used a structure inspired by chemical models of coal [62]. A small fragment of this is shown in fig. 7, and fig. 8 displays the frequency spectrum of its elecric dipole fluctuations after deposition of $\sim$4.5 eV in one C-H bond. If the same energy were thermalized, the peak temperature reached by the particle would be $\sim$ 300K and the ratio [3.3]/[11.3] would fall below 1$\%$ !

Experiments such as that of Williams and Leone [21] and Wagner et al. [15] seem to show that randomization in PAHs is also slow after absorption of a UV photon, so that IR \emph{photo}luminescence is observed. This process could therefore coexist with chemiluminescence. However, it probably suffers from the undesirable loss of part of the deposited energy to vis/UV fluorescence and electronic continuum emission; neither of these affects chemiluminescence since the chemical excitation considered here leaves the system in the electronic ground state.

\begin{figure}
%\vspace{8cm}
%\begin{center}
%\scalebox{0.5}[0.5]{
\resizebox{\hsize}{!}{\includegraphics{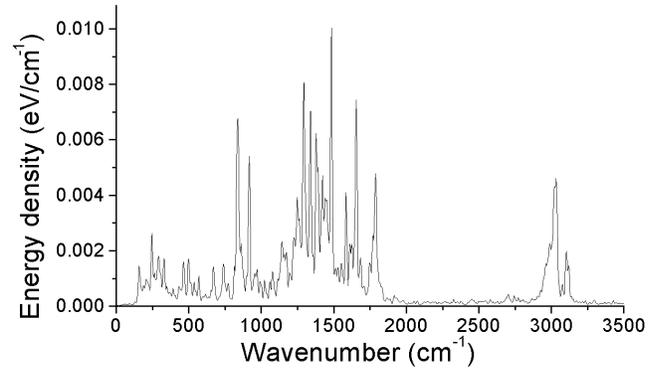}}
%\end{center}
\caption[ ]{The spectrum of the fragment of dust grain modelled in Fig. 7, 12.2 ps after capture of an H atom; smoothed to a resolution of 13 cm$^{-1}$.}
\end{figure}

\vskip1cm
\section{Conclusion}
One is overwhelmed by the immense diversity of both UIB and coal/kerogen IR spectra. Some relief is provided by 1) the sparsness of the prominent bands, 2) their relative stability in wavelength, 3) the encouraging fits attempted between observations and models over most of the UIBs and extended to the UV/vis and far IR ranges, 4) the vast leeway for model tailoring, offered by the disordered structure of coals and kerogens, as opposed to the very restrictive aromaticity of PAHs, 5) the credibility of the model material, because it is so easily found in nature and essentially made of only the four most abundant elements in space (after He), and 6) the fact that this material has already been so thoroughly studied in the laboratory that even its diversity is mostly understood.

The issue of the excitation mechanism is intimately linked to the issue of modelling the observed spectra. By invoking chemiluminescence on large particles, one avoids high dust temperatures and the attendent puzzles of varying band wavelengths and widths which are not observed in the sky.

Among the issues that seem now to deserve early investigation are

- the assignment of the ``lone" 11.3-$\mu$m feature (as opposed to the mono-, duo-, etc. out-of-plane C-H bends);

- the assignment of the 12.7 $\mu$m feature;

-the decreasing contrast of UIBs with increasing distance from central star, in certain objects (see [42]);

-the experimental demonstration of IR photo- and/or chemiluminescence on \emph{large isolated} fragments of coal or kerogen and measurement of their efficiencies;

-the need for a library of complete celestial IR spectra ($\sim$2 to $\sim$16 $\mu$m), each associated with a known environment, differing from the others by the intensity of the radiation field or the nature of the object : RN, PPN, PN, PDR, GC, etc.; this would help understand, by analogy with coals and kerogens, the evolution of IS dust from inception in the envelopes of AGB stars to its consumption in protostars;

-the measurement of the visible absorption spectrum of kerogen, to be compared with the DIBs.

The coal/kerogen model can set a frame for the discussion of these and other issues. However, keeping in mind that these terrestrial materials were forged under special physical and chemical conditions, the model will no doubt have to be amended according to observations, as the debate goes on.

\begin{acknowledgements}
I am indebted to J. Conard, J.-N. Rouzaud (CNRS/Orleans), and M. Cauchetier, O. Guillois, I. Nenner, C. Reynaud (CEA/DRECAM, Saclay) for a fruitful collaboration, and to many others for enlightening discussions.
\end{acknowledgements}

References
\vskip0.5cm
1. Cox P. and Kessler M. (Eds) 1999, The Universe as seen by ISO, ESA SP-427, Noordwijk, NL

2. Murakami H. et al. 1996, PASJ 48, L41

3. Tokunaga A. 1997, in ASP Conf. Ser. 124, 149

4. Boulanger F. et al. 1998, in ASP Conf. Ser. 132, 15

5. Sellgren K. 2001, Spectrochim. Acta 57A, 627

6. Donn B. et al. 1989, in Interstellar Dust, IAU 135, Allamandola and Tielens A. (Eds), 181

7. Papoular R. et al. 1996, Astron. Astrophys. 315, 222

8. Papoular R. et al. 1989, Astron. Astrophys. 217, 204

9. Guillois O. et al. 1996, Ap J 467, 810

10. Papoular R. 2001, Astron. Astrophys. 378, 597

11. Papoular R. 2000, Astron. Astrophys. 359, 397

12. Behar F. and Vandenbroucke M. 1986, Rev. Inst. Fr. Petrole 41(2), 173

13. Robin P. and Rouxhet P. 1976, Rev. Inst. Fr. Petrole 31(6), 955

14. Durand B. 1980, Kerogen, Technip, Paris

15. Wagner D. et al. 2000, Ap J 545, 854

16. Cody G. Jr  et al. 1990, Mat. Res. Soc. Symp. Proc. vol. 195, p.559

17. Duley W. and Williams D. 1984, Interstellar Chemistry, Acad. Pr., London

18. Hollis J. et al. 2002, Ap J 571, L59

19. Papoular R. 2000, Astron. Astrophys. 362, L9

20. Papoular R. 1999, Astron. Astrophys. 346, 219

21. Williams R. and Leone S. 1995, Ap J 443, 675

22. Verstraete L. et al. 2001, Astron. Astrophys. 372, 981

23. Verstraete L. 2001, in ISO: beyond the peaks, ESA SP-456, 319

24. Boulanger F. et al. 2000, ISO Beyond Point Sources, ESA SP-455, 91

25. Elliott S. 1990, Physics of amorphous materials, Longman, New York

26. Magazzu A. and Strazzulla 1992, Astron. Astrophys. 263, 281

27. Hylands A. and McGregor P. 1988, in Interstellar Dust, IAU 135, Contributed Papers, NASA CP 3036, 101

28. Pendleton Y., Sandford S., Allamandola L. et al., 1994, ApJ 437, 683

29. Joblin C., Tielens A., Allamandola L. and Geballe T., 1996, ApJ 458, 610

30. Geballe T., Tielens A., Kwok S. and Hrivnak B., 1992, ApJ Lett., 387, L29

31. Sellgren K., 1984, ApJ 277, 623

32. Guillois O., 1996, Thesis, Paris-Sud University

33. Colthup. N., Daly L. and Wiberley S. 1990, Introduction to Infrared and Raman Spectroscopy, Acad. Pr., Boston

34. Dischler B. 1987, E-MRS Meeting, XVII, Editions de Physique, Paris, 189

35. Hony S. et al. 2001, Astron. Astrophys. 370, 1030

36. Davenas J. 1993, Solid State Phenomena 30/31, 317

37. Robertson D., Brenner D. and White C. 1992, J. Phys. Chem. 96, 6133 

38. Huffman D. 1991, Phys. Today 44, 22

39. Abergel A. et al. 2002, Astron. Astrophys. 389, 239

40. Cesarsky D. et al. 2000, Astron. Astrophys. 354, L87

41. Miville-Deschenes et al. 2002, Astron. Astrophys. 381, 209

42. Cesarsky D. et al. 2000, Astron. Astrophys. 358, 708

43. Cook D. and Saykally R. 1998, Ap J 493, 793

44. Pendleton Y. and Allamandola L. 2002, Ap J Suppl. 138, 75

45. Birks J. 1970, Photophysics of aromatic molecules, Wiley, Boston

46. Snow T. 2001, Spectrochimica Acta 57A, 615

47. Cohen M. et al. 1986, ApJ 302, 737

48. Chan K.-W. et al. 2001, Ap J 546, 273

49. Haas M. et al. 2002, Astron. Astrophys. 385, L23

50. Vermeij R. et al. 2002, Astron. Astrophys. 382, 1042

51. Waters L. et al. 1989, Astron. Astrophys. 211, 208

52. Peeters E. et al. 2002, Astron. Astrophys. 390, 1089

53. Leger A. and Puget J.-L. 1989, An. Rev. Astron. Astrophys. 27, 161

54. Draine B. and Li A. 2001, Ap J 551, 807

55. Sellgren K. et al. 1990, Ap J 349, 120

56. Graham J. et al. 1993, Astron. J. 105, 250

57. Guillois O. et al. 1998, Chemistry and Physics of Molecules and Grains in Space, Faraday Disc. 109, 335

58. Cashion J. and Polanyi J. 1959, J. Chem. Phys. 30, 1097L

59. Papoular R. 2001, Spectrochimica Acta, 57A, 947

60. Papoular R. 2002, J. Phys. B, 35, 3741

61. Pagani L. et al. 1999, Astron. Astrophys. 351, 447

62. Speight J. 1994, Appl. Spectr. Rev. 29, 117

63. Bregman J. et al. 1994, Ap J 423, 326

64. Crete E. et al. 1999, Astron. Astrophys. 352, 277

65. Habart E. et al. 2000, in ISO Beyond the Peaks, ESA SP-456, Noordwijk, NL, p. 103

\end{document}